\documentclass[preprint,aps,nofootinbib]{revtex4}
\usepackage{amsfonts,amsmath,amssymb,amsthm}
\usepackage{latexsym}
\usepackage{bbm,bm}
\usepackage{graphicx}

\newcommand{\ket}[1]{\lvert #1 \rangle}
\newcommand{\bra}[1]{\langle #1 \lvert}
\newcommand{\beq}{\begin{equation}}
\newcommand{\eeq}{\end{equation}}
\newcommand{\beqs}{\begin{eqnarray}}
\newcommand{\eeqs}{\end{eqnarray}}
\begin{document}

\title{Quantum Discord and Quantum Entanglement in the Background of an Asymptotically Flat Static Black Holes}

\author{Eylee Jung$^1$ Mi-Ra Hwang$^2$, and DaeKil Park$^{1,2}$}

\affiliation{$^1$Department of Physics, Kyungnam University, ChangWon
                  631-701, Korea    \\
             $^2$Department of Electronic Engineering, Kyungnam      University, ChangWon
                 631-701, Korea                                         
             }

\begin{abstract}
The quantum discord and tripartite entanglement are discussed in the presence of an asymptotically flat static black holes.
The total correlation, quantum discord and classical correlation exhibit decreasing behavior with increasing Hawking temperature. It is shown 
that the classical correlation is less than the quantum discord in the full range of Hawking temperature. The tripartite
entanglements for Greenberger-Horne-Zeilinger and W-states also exhibit decreasing behavior with increasing Hawking temperature. 
At the infinite limit of Hawking temperature the tripartite entanglements for Greenberger-Horne-Zeilinger and W-states reduce to $52 \%$ and $33 \%$ 
of the corresponding values in the flat space limit, respectively.
\end{abstract}

\maketitle

\section{Introduction}
Recently much attention is paid to the quantum information theories in the relativistic framework\cite{peres1,peres2,peres3,inertial1,inertial2,inertial3,inertial4,noninertial1,noninertial2,noninertial3,
noninertial4,noninertial5,noninertial6,noninertial7,noninertial8,noninertial9,noninertial10,noninertial11,noninertial12}. 
The most remarkable fact in the 
inertial frames is the fact that entanglement of given multipartite quantum state is conserved although the entanglement between 
some degrees of freedom can be transfered to others\cite{inertial1,inertial2,inertial3,inertial4}. In non-inertial frames, however, the 
entanglement is in general degraded, which implies that the quantum correlation between rest and accelerating observers is reduced 
more and more with increasing the acceleration\cite{noninertial1}. The main reason for the reduction of the quantum correlation is 
that the accelerating observer located in one Rindler wedge loses an information arising from the other Rindler wedge due to the 
causally disconnected nature between the wedges. This means that some quantum information is leaked into other causally disconnected
Rindler space, which makes the reduction of the quantum correlation. In fact, this is a main scenario of the well-known Unruh
effect\cite{unruh1,unruh2}. Recently, this Unruh-type decoherence effect beyond the single-mode approximation is discussed in the 
context of the quantum information theories\cite{beyond}.

More recently, the quantum entanglement in the black hole background is examined\cite{black1,black2}. Especially, in Ref.\cite{black1} the 
Hawking temperature-dependence of the bipartite entanglement is studied in the arbitrary spherically symmetric and asymptotically flat
black hole background. The purpose of this paper is to explore the quantum discord\cite{zurek02,vedral01} and the tripartite entanglement
in the same black hole background.

\section{spacetime background}
Throughout this paper we use $G = c = \hbar = k_B = 1$. The metric we consider in this paper is 
\begin{equation}
\label{metric1}
ds^2 = f(r) dt^2 - \frac{1}{h(r)} dr^2 - R^2 (r) \left( d\theta^2 + \sin^2 \theta d\varphi^2 \right),
\end{equation}
where the functions $f(r)$, $h(r)$, and $R(r)$ satisfy $f(\infty) = h(\infty) = 1$, $R(\infty) = r$, and $f(r_H) = h(r_H) = 0$.
Thus, the line element (\ref{metric1}) includes the various black holes such as Schwarzschild and Reissner-Nordstr\"om black holes.
The Hawking temperature in this metric is $T_H = \kappa / 2 \pi$, where $\kappa$ is a surface gravity defined as 
$\kappa = \sqrt{f'(r_H) h'(r_H)} / 2$. 

As shown in Ref. \cite{black1} one can consider three-different vacuum states $\ket{0}_{in}$, $\ket{0}_{out}$, and $\ket{0}_K$ in 
this background. First two vacuum states are the Fock vacua inside and outside horizon, respectively, and the last one is the 
Kruskal vacuum outside the event horizon. The interrelation between these vacua is 
\begin{equation}
\label{inter-0}
\ket{0}_K = \sqrt{1 - e^{-\omega / T_H}} \sum_{n=0}^{\infty} e^{-n \omega / 2 T_H} \ket{n}_{in} \otimes \ket{n}_{out},
\end{equation}
where $\ket{n}_{in}$ and $\ket{n}_{out}$ are $n$-particle states constructed from $\ket{0}_{in}$ and $\ket{0}_{out}$ by operating 
the corresponding creation operators $n$ times, and $\omega$ is a frequency of the scalar field. Applying the creation operator of the 
Kruskal spacetime in Eq. (\ref{inter-0}) and using the Bogoliubov coefficients, one can construct $\ket{1}_K$, whose expression is 
\begin{equation}
\label{inter-1}
\ket{1}_K = \left(1 - e^{-\omega / T_H} \right) \sum_{n=0}^{\infty} \sqrt{n+1} e^{-n \omega / 2 T_H} \ket{n}_{in} \otimes \ket{n+1}_{out}.
\end{equation}

\section{Quantum Discord}
Quantum discord\cite{zurek02,vedral01} is a measure for {\it quantumness} of given bipartite quantum state. Usually these two parties
consist of system and corresponding apparatus. In this paper, however, we will call these parties as Alice and Bob. We will examine
in this section how the quantum discord is changed in the presence of the black hole (\ref{metric1}).

We assume that initially Alice and Bob share a entangled state
\begin{equation}
\label{initial}
\ket{\psi}_{AB} = \frac{1}{\sqrt{2}} \bigg(\ket{1}_A \ket{0}_B + \ket{0}_A \ket{1}_B \bigg)
\end{equation}
in the asymptotic region. After sharing, Bob moves to the near-horizon region with his own particle detector while Alice stays in the asymptotic region. 
Since, then, the Bob's detector registers only thermally excited particles due to the Hawking effect,
Bob's state can be represented by tensor product of the $in-$ and $out-$ states.
However, since the inside region of the black hole is causally disconnected from Alice and Bob, we have to take a partial trace over
$in-$state. Then, the state between Alice and Bob becomes a mixed state whose density matrix becomes
\begin{equation}
\label{density-1}
\rho_{AB} = \frac{1}{2} \ket{0}_A \bra{0} \otimes M_{00} + \frac{1}{2} \ket{1}_A \bra{1} \otimes M_{11} +
            \frac{1}{2} \ket{0}_A \bra{1} \otimes M_{01} + \frac{1}{2} \ket{1}_A \bra{0} \otimes M_{10}
\end{equation}
where 
\begin{eqnarray}
\label{density-2}
& &M_{00} = (1 - e^{-\omega / T_H} ) \sum_{n=0}^{\infty} e^{-n \omega / T_H} \ket{n}\bra{n}              \nonumber  \\
& &M_{11} = (1 - e^{-\omega / T_H} )^2 \sum_{n=0}^{\infty} (n+1) e^{-n \omega / T_H} \ket{n+1}\bra{n+1}    \\
& &M_{01} = (1 - e^{-\omega / T_H} )^{3/2} \sum_{n=0}^{\infty} \sqrt{n+1} e^{-n \omega / T_H} \ket{n}\bra{n+1}  \nonumber  \\
& &M_{10} = (1 - e^{-\omega / T_H} )^{3/2} \sum_{n=0}^{\infty} \sqrt{n+1} e^{-n \omega / T_H} \ket{n+1}\bra{n}.          \nonumber
\end{eqnarray}
It is worthwhile noting $\mbox{Tr}_B M_{00} = \mbox{Tr}_B M_{11} = 1$ and $\mbox{Tr}_B M_{01} = \mbox{Tr}_B M_{10} = 0$.

Now, we discuss the quantum discord. We assume that Alice performs a projective measurement with a complete set of the 
measurement operators $\{ \Pi_j^A \}$. The usual mutual information between Alice and Bob is 
\begin{equation}
\label{mutual1}
I (A:B) = S(A) + S(B) - S(A,B)
\end{equation}
where $S$ denotes the von Neumann entropy $S(\rho) = \mbox{Tr} (\rho \log \rho)$. In our paper, all logarithms are taken to base $2$.
The classical analogue of Eq. (\ref{mutual1}) is $I_{cl} (A:B) = H(A) + H(B) - H(A,B)$, where $H$ denotes the Shannon entropy. In classical 
information theories different representation of the mutual information is $I_{cl} (A:B) = H(A) - H(A|B) = H(B) - H(B|A)$, where 
$H(X|Y)$ is the conditional entropy of $X$ given $Y$. The quantum analogue of this representation\cite{zurek02} is 
\begin{equation}
\label{mutual2}
J (A:B)_{\left\{ \Pi_j^A \right\}} = S(B) - \sum_j p_j S(B | \Pi_j^A ),
\end{equation}
where $\left\{ \Pi_j^A \right\}$ denotes a complete set of the measurement operators prepared by the party $A$
and $S(B | \Pi_j^A )$ is a von Neumann entropy of the party $B$ after the party $A$ has a measurement outcome $j$. Of course, 
$p_j$ is a probability for getting outcome $j$ in the quantum measurement. Usual quantum mechanical postulates\cite{postulate} imply 
\begin{equation}
\label{mutual3}
p_j = \mbox{Tr}_{A,B} ( \Pi_j^A \rho_{AB} \Pi_j^A )            \hspace{1.0cm}
S(B | \Pi_j^A ) = S \bigg( \rho \left(B | \Pi_j^A \right) \bigg)
\end{equation}
where $\rho \left(B | \Pi_j^A \right) = \mbox{Tr}_A ( \Pi_j^A \rho_{AB} \Pi_j^A ) / p_j$. Unlike $I (A:B)$, therefore, $J (A:B)$ is dependent
on the complete set of the measurement operators. The quantum discord is defined as 
\begin{equation}
\label{discord1}
{\cal D} (A:B) = \min \left[ I(A:B) - J(A:B)\right] = \min \left[S(A) - S(A,B) + \sum_j p_j S(B | \Pi_j^A ) \right],
\end{equation}
where minimum is taken over all possible complete set of the measurement operators\footnote{Although authors in Ref. \cite{zurek02} considers
the projective measurement, authors in Ref. \cite{vedral01} considers the general measurement including POVM. Thus, the latter is the lower 
bound of the former.}.

Now, we would like to compute the quantum discord in the black hole background. From Eq. (\ref{density-1}) it is easy to show that 
$\rho_A \equiv \mbox{Tr}_B \rho_{AB}$ is a completely mixed state and 
\begin{equation}
\label{Neumann-A}
S(A) = 1.
\end{equation}
Also it is easy to show
\begin{eqnarray}
\label{Neumann-AB}
& &S(A,B) = -\sum_{n=0}^{\infty} \Lambda_n \log \Lambda_n                            \\   \nonumber
& &\Lambda_n = \frac{1}{2} e^{-n\omega / T_H} \left(1 - e^{-\omega / T_H} \right) \bigg[1 + (n+1) \left(1 - e^{-\omega / T_H} \right) \bigg].
\end{eqnarray}

%%%%%%%%%%%%%%%%%%%%%%%%%%%%%%%%%%%%%%%%%%%%%%%%%%%%%%%%%
\begin{figure}[ht!]
\begin{center}
\includegraphics[height=8cm]{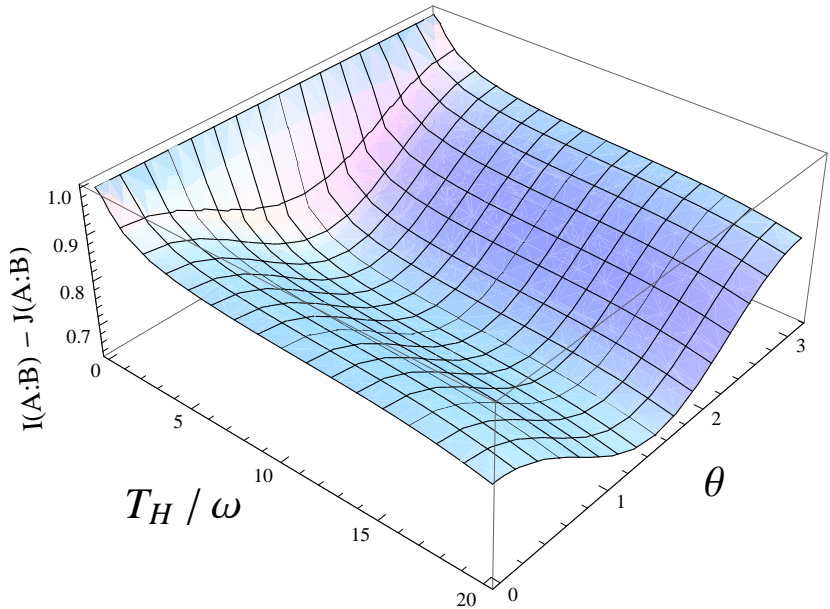}
\caption[fig1]{The $\theta$- and Hawking temperature-dependence of $I(A:B) - J(A:B)$. Minimum of $I(A:B) - J(A:B)$ occurs at 
$\theta = \pi/2$ in the full range of Hawking temperature.}
\end{center}
\end{figure}
%%%%%%%%%%%%%%%%%%%%%%%%%%%%%%%%%%%%%%%%%%%%%%%%%%%%%%%%%%%

Now, we introduce the complete set of the projective measurement operators $\left\{ \Pi_1^A, \Pi_2^A \right\}$ with
\begin{equation}
\label{projective-1}
\Pi_1^A = \frac{I_2 + {\bm x} \cdot {\bm \sigma}}{2}     \hspace{1.0cm}
\Pi_2^A = \frac{I_2 - {\bm x} \cdot {\bm \sigma}}{2}.
\end{equation}
In Eq. (\ref{projective-1}) ${\bm \sigma}$ denotes the Pauli matrix and $x_1^2 + x_2^2 + x_3^2 = 1$. Then, it is straightforward to show 
$p_1 = p_2 = 1/2$ and 
\begin{eqnarray}
\label{projective-2}
& &\rho(B|\Pi_1^A) = \frac{1}{2} \bigg[ (1 + x_3) M_{00} + (1 - x_3) M_{11} + (x_1 + i x_2) M_{01} + (x_1 - i x_2) M_{10} \bigg]  \\   \nonumber
& &\rho(B|\Pi_2^A) = \frac{1}{2} \bigg[ (1 - x_3) M_{00} + (1 + x_3) M_{11} - (x_1 + i x_2) M_{01} - (x_1 - i x_2) M_{10} \bigg].
\end{eqnarray}
Since it is impossible to compute the eigenvalues of the $\rho(B|\Pi_j^A) \hspace{.2cm} (j=1,2)$, we should compute 
$S \bigg( \rho \left(B | \Pi_j^A \right) \bigg)$ numerically. One can perform this numerical calculation with
parametrizing $x_1 = \sin \theta \cos \phi$,
$x_2 = \sin \theta \sin \phi$ and $x_3 = \cos \theta$. Then, it is possible to show that the eigenvalues of $\rho(B|\Pi_j^A)$ are independent of $\phi$.

The $(T_H / \omega, \theta)$-dependence of $I(A:B) - J(A:B)$ is plotted in Fig. 1. As this figure exhibits, the minimum is occurred at $\theta = \pi/2$.
Therefore, the quantum discord ${\cal D} (A:B)$ is obtained from $I(A:B) - J(A:B)$ by letting $\theta = \pi/2$. If we assume that the total 
correlation is a mutual information $I (A:B)$, it is possible to compute the classical correlation ${\cal C} (A:B)$ by
\begin{equation}
\label{classical-1}
{\cal C} (A:B) = I (A:B) - {\cal D} (A:B).
\end{equation}

%%%%%%%%%%%%%%%%%%%%%%%%%%%%%%%%%%%%%%%%%%%%%%%%%%%%%%%%%
\begin{figure}[ht!]
\begin{center}
\includegraphics[height=8cm]{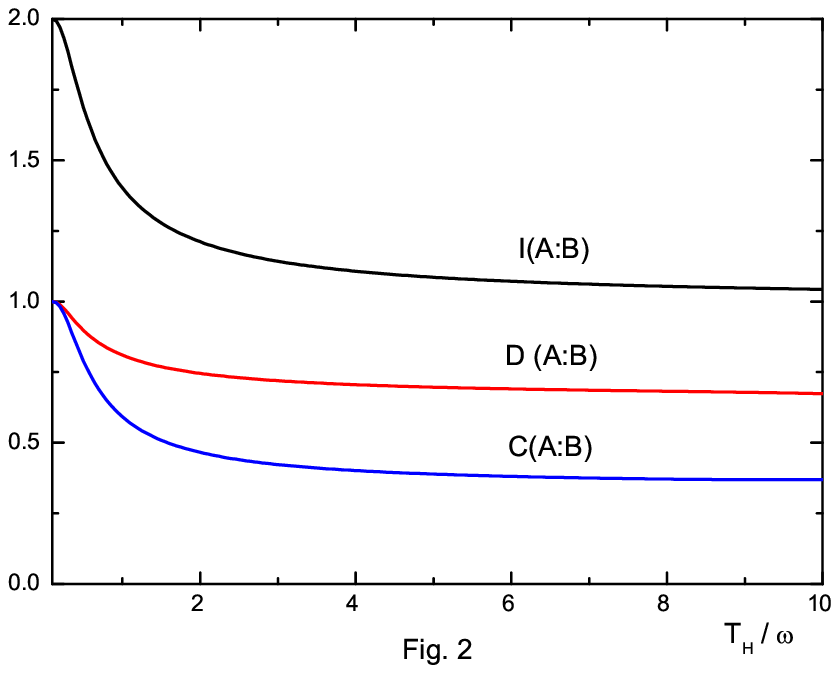}
\caption[fig2]{The Hawking temperature-dependence of total correlation, quantum discord, and classical correlation. All correlations show
a decreasing behavior with increasing the temperature and reduce to $50 \%$, $60 \%$, and $40 \%$ of the corresponding values in the flat space limit
at $T_H = \infty$.}
\end{center}
\end{figure}
%%%%%%%%%%%%%%%%%%%%%%%%%%%%%%%%%%%%%%%%%%%%%%%%%%%%%%%%%%%

In Fig. 2 we plot the Hawking temperature-dependence of the total correlation, quantum discord, and classical correlation. 
As Fig. 2 shows, all correlations exhibit a decreasing behavior with increasing $T_H$. In the $T_H \rightarrow 0$ limit all correlations approach to
the values in the absence of the black hole. In the opposite limit, i.e. $T_H \rightarrow \infty$, $I (A:B)$, ${\cal D} (A:B)$, and ${\cal C} (A:B)$
approach to $1.0$, $0.6$, and $0.4$, respectively. The remarkable fact is that the classical correlation is less than the quantum discord
in the full range of Hawking temperature. Similar behavior was derived when the classical correlation and quantum discord sharing of Dirac
field are discussed in the non-inertial frame\cite{noninertial5}. In next section we will discuss on the tripartite entanglement in the 
presence of the black hole (\ref{metric1}).

\section{Tripartite Entanglement Degradation}

The most well-known measure for the tripartite entanglement 
is a three-tangle\cite{ckw}. Since, however, the three-tangle is not defined in the qudit system, we cannot use it because of Eq. (\ref{inter-0}) and
Eq. (\ref{inter-1}). Thus, instead of the three-tangle, we will use $\pi$-tangle\cite{ou07} in this paper as a measure of the tripartite 
entanglement. 

\subsection{Greenberger-Horne-Zeilinger state}

Let Alice, Bob, and Charlie share the Greenberger-Horne-Zeilinger (GHZ) state 
\begin{equation}
\label{GHZ-1}
\ket{GHZ}_{ABC} = \frac{1}{\sqrt{2}} \left[ \ket{000} + \ket{111} \right]_{ABC}
\end{equation}
in the asymptotic flat region. If Charlie moves to the near-horizon region with his own particle detector, Eq. (\ref{inter-0}) and Eq. (\ref{inter-1}) with tracing over the 
Charlie's $in$-state imply
\begin{eqnarray}
\label{GHZ-2}
& & \ket{GHZ}_{ABC}                                                                                          \\   \nonumber
&\rightarrow& \rho_{ABC} = \frac{1}{2} \sum_{n=0}^{\infty} e^{-n \omega / T_H}                                
 \Bigg[ \nu \ket{00n}\bra{00n}                                                   
+ \nu^2   (n+1)  \ket{11(n+1)}\bra{11(n+1)}                                                  \\   \nonumber
& &   \hspace{3.0cm}           + \nu^{3/2}  \sqrt{n+1}  \bigg\{ \ket{00n}\bra{11(n+1)} + \ket{11(n+1)}\bra{00n} \bigg\}
        \Bigg],
\end{eqnarray}
where $\nu = 1 - e^{-\omega / T_H}$.
Since the Charlie's $out$-state is a qudit state, it is impossible to compute the genuine tripartite entanglement measure called the 
three-tangle\cite{ckw}. As we commented before, therefore, we choose the $\pi$-tangle\cite{ou07} as a tripartite measure defined as 
\begin{equation}
\label{pi-t1}
\pi = \frac{1}{3} (\pi_A + \pi_B + \pi_C )
\end{equation}
due to more tractable computation. In Eq. (\ref{pi-t1}) $\pi_A$, $\pi_B$, and $\pi_C$ are defined by
\begin{equation}
\label{pi-t2}
\pi_A = {\cal N}^2_{A(BC)} - {\cal N}^2_{AB} - {\cal N}^2_{AC}             \hspace{.5cm}
\pi_B = {\cal N}^2_{B(AC)} - {\cal N}^2_{AB} - {\cal N}^2_{BC}             \hspace{.5cm}
\pi_C = {\cal N}^2_{C(AB)} - {\cal N}^2_{AC} - {\cal N}^2_{BC},             
\end{equation}
where ${\cal N}_{\alpha(\beta\gamma)} = ||\rho_{ABC}^{T_{\alpha}}|| - 1$ and ${\cal N}_{\alpha\beta} = ||(\mbox{Tr}_{\gamma}\rho_{ABC}^{T_{\alpha}}||-1$
with $T_{\alpha}$ being a partial transposition over $\alpha$-state and $||A||= \mbox{Tr} \sqrt{A A^{\dagger}}$. It is easy to show $\pi_{GHZ} = 1$
in the absence of the black hole background.

Now, let us compute the one-tangle ${\cal N}_{A(BC)}$. From Eq. (\ref{GHZ-2}) it is easy to show that 
$\left(\rho_{ABC}^{T_A}\right) \left(\rho_{ABC}^{T_A}\right)^{\dagger}$ is a diagonal. Therefore, the eigenvalues of 
$\left(\rho_{ABC}^{T_A}\right) \left(\rho_{ABC}^{T_A}\right)^{\dagger}$ can be computable easily. Since $||\rho_{ABC}^{T_A}||$ is a 
sum of square root of the eigenvalues, one can derive ${\cal N}_{A(BC)}$, whose final expression is 
\begin{equation}
\label{one-t1}
{\cal N}_{A(BC)} = \nu^{3/2} e^{\omega / T_H}  Li_{-1/2} \left(e^{-\omega/T_H}\right),
\end{equation}
where $Li_n (z)$ is a polylogarithm function defined as 
\begin{equation}
\label{poly-log}
Li_n (z) \equiv \sum_{k=1}^{\infty} \frac{z^k}{k^n} = \frac{z}{1^n} + \frac{z^2}{2^n} + \frac{z^3}{3^n} + \cdots.
\end{equation}
Using a property of the polylogarithm function one can show that ${\cal N}_{A(BC)}$ approaches to $\sqrt{\pi} / 2$ when $T_H \rightarrow \infty$.
From a symmetry of the GHZ state it is also easy to show ${\cal N}_{B(AC)} = {\cal N}_{A(BC)}$. 

Now, let us compute the last one-tangle ${\cal N}_{C(AB)}$. Since $\left(\rho_{ABC}^{T_C}\right) \left(\rho_{ABC}^{T_C}\right)^{\dagger}$
becomes
\begin{equation}
\label{one-t2}
\left(\rho_{ABC}^{T_C}\right) \left(\rho_{ABC}^{T_C}\right)^{\dagger} = D + F,
\end{equation}
where $D$ and $F$ are 
\begin{eqnarray}
\label{one-t3} 
& &D = \frac{1}{4} \sum_{n=0}^{\infty} e^{-2 n \omega / T_H}                                
 \Bigg[ \nu^2 \ket{00n}\bra{00n}                                                   
+ \nu^4   (n+1)^2  \ket{11(n+1)}\bra{11(n+1)}                                                  \\   \nonumber
& &   \hspace{4.0cm}           + \nu^{3}  (n+1)  \bigg\{ \ket{00(n+1)}\bra{00(n+1)} + \ket{11n}\bra{11n} \bigg\}
        \Bigg]                                                                                                       \\   \nonumber
& &F = \frac{1}{4} \sum_{n=0}^{\infty} e^{-(2n+1) \omega / T_H}
  \Bigg[ \nu^{5/2} \sqrt{n+1} \bigg\{ \ket{11n}\bra{00(n+1)} + \ket{00(n+1)}\bra{11n} \bigg\}                          \\    \nonumber
& &  \hspace{2.0cm}  
+ \nu^{7/2} (n+1) \sqrt{n+2} \bigg\{ \ket{11(n+1)}\bra{00(n+2)} + \ket{00(n+2)}\bra{11(n+1)} \bigg\} 
   \Bigg],
\end{eqnarray}
the off-diagonal part $F$ makes it difficult to compute the eigenvalues of $\left(\rho_{ABC}^{T_C}\right) \left(\rho_{ABC}^{T_C}\right)^{\dagger}$.
However, one can make $\left(\rho_{ABC}^{T_C}\right) \left(\rho_{ABC}^{T_C}\right)^{\dagger}$ block-diagonal by ordering the basis
as $\left\{ \ket{000}, \ket{110}, \ket{001}, \ket{111}, \ket{002}, \ket{112}, \cdots \right\}$. Thus, one can compute the 
eigenvalues of $\left(\rho_{ABC}^{T_C}\right) \left(\rho_{ABC}^{T_C}\right)^{\dagger}$ analytically, which are 
$\left\{ \nu^2 / 4, \Lambda_n^{\pm} \bigg|_{n=0, 1, 2, \cdots} \right\}$. Here, $\Lambda_n^{\pm}$ are eigenvalues of each block given by
\begin{equation}
\label{one-t4}
\Lambda_n^{\pm} = \frac{\nu^2}{8} e^{-2n\omega / T_H} \left[ (\mu_n^2 + 2 \nu) \pm \mu_n \sqrt{\mu_n^2 + 4 \nu} \right],
\end{equation}
where $\mu_n = n e^{\omega / T_H} \nu + e^{-\omega / T_H}$. Therefore, ${\cal N}_{C(AB)}$ reduces to 
\begin{equation}
\label{one-t5}
{\cal N}_{C(AB)} = \frac{\nu}{2} + \sum_{n=0}^{\infty} \left(\sqrt{\Lambda_n^+} + \sqrt{\Lambda_n^-} \right) - 1.
\end{equation}
Finally, one can show that all two tangles ${\cal N}_{AB}$, ${\cal N}_{AC}$, and ${\cal N}_{BC}$ are identically zero.

%%%%%%%%%%%%%%%%%%%%%%%%%%%%%%%%%%%%%%%%%%%%%%%%%%%%%%%%%
\begin{figure}[ht!]
\begin{center}
\includegraphics[height=8cm]{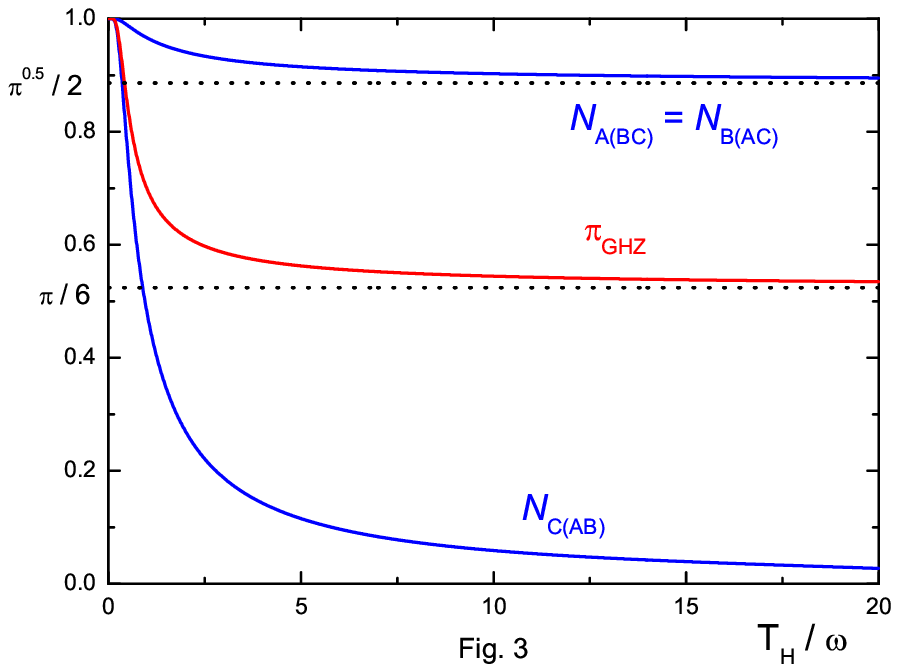}
\caption[fig3]{The Hawking temperature-dependence of one tangles and $\pi_{GHZ}$. The $\pi$-tangle decreases with increasing 
$T_H$, and eventually reduces to $\pi / 6 \sim 0.524$ at $T_H = \infty$.}
\end{center}
\end{figure}
%%%%%%%%%%%%%%%%%%%%%%%%%%%%%%%%%%%%%%%%%%%%%%%%%%%%%%%%%%%

The one-tangles and $\pi$-tangle are plotted in Fig. 3 as a function of Hawking temperature. As this figure shows, the $\pi$-tangle decreases
with increasing the Hawking temperature, and eventually reduces to $\pi /6 \sim 0.524$ at $T_H \rightarrow \infty$. At $T_H = 0$ the
$\pi$-tangle exactly coincides with that in the absence of the black hole. Thus, as expected, the tripartite entanglement is degraded when Charlie 
moves to the near-horizon region from the asymptotic region with his own particle detector. 

\subsection{W state}

Let Alice, Bob, and Charlie share the W-state\cite{dur00-1} 
\begin{equation}
\label{W-1}
\ket{W}_{ABC} = \frac{1}{\sqrt{3}} \left[ \ket{001} + \ket{010} + \ket{100} \right]_{ABC}
\end{equation}
in the asymptotic flat region. It is easy to show that the $\pi$-tangle for the W-state is $\pi_W = 4 (\sqrt{5} - 1) / 9 \sim 0.55$ in the flat space limit.

By following similar calculation to the case of GHZ state, 
one can compute the Hawking temperature-dependence of $\pi_W$ in the presence of the black hole
background. We do not want to repeat the computational procedure again in this paper. Instead, we present Fig. 4, which shows one-tangles, 
two-tangles, and $\pi_W$ as a function of Hawking temperature. In Fig. 4 (a) we plot the one- and two-tangles. All tangles exhibit decreasing
behavior with increasing Hawking temperature except ${\cal N}_{AB}$, which is independent of $T_H$. At $T_H \rightarrow \infty$ ${\cal N}_{A(BC)}$ 
and ${\cal N}_{B(AC)}$ approaches to $0.659$ while ${\cal N}_{C(AB)}$ has a vanishing limit. The remarkable fact is that the two-tangles
${\cal N}_{AC}$ and ${\cal N}_{BC}$ becomes abruptly zero in the region $T_H > 1.45 \omega$. This reminds us of the concurrence, one of the 
bipartite entanglement measure. In Fig. 4 (b) we plot $\pi_W$ as a function of $T_H$. 
At $T_H = 0$ $\pi_W$ in the flat space is recovered. However, it exhibits a decreasing behavior with increasing $T_H$, and eventually reduces to 
$0.18$ at $T_H = \infty$ limit.

%%%%%%%%%%%%%%%%%%%%%%%%%%%%%%%%%%%%%%%%%%%%%%%%%%%%%%%%%
\begin{figure}[ht!]
\begin{center}
\includegraphics[height=6.0cm]{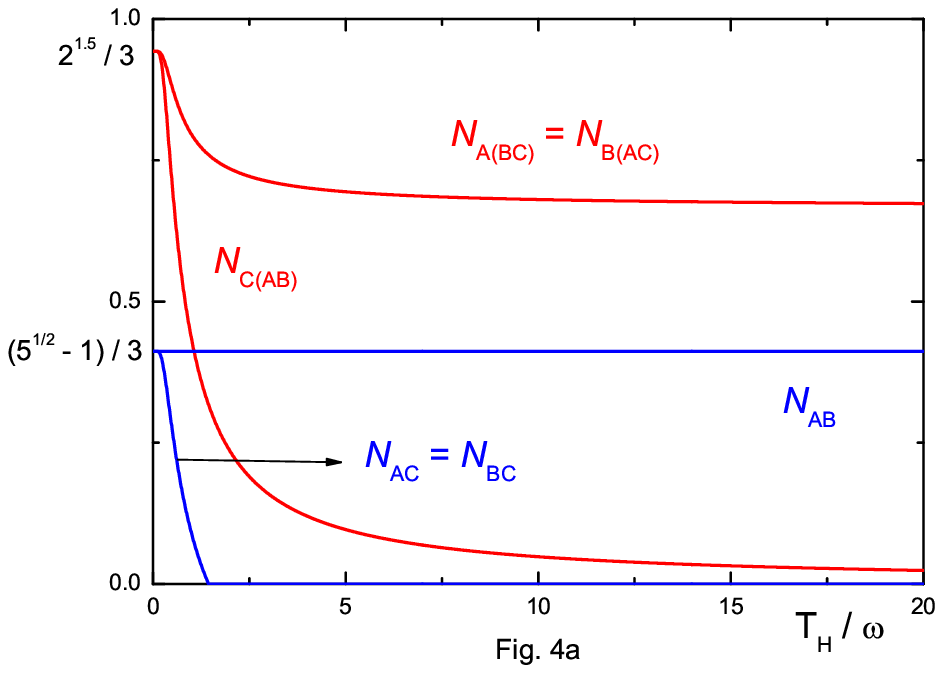}
\includegraphics[height=6.0cm]{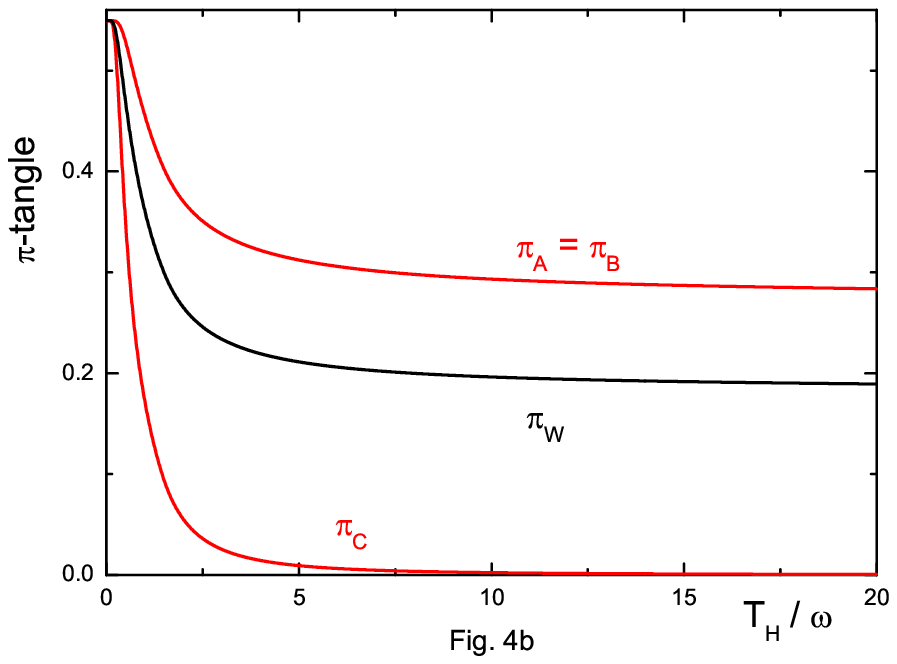}
\caption[fig4]{(a) The Hawking temperature-dependence of one- and two-tangles. (b) The Hawking temperature-dependence of $\pi_W$. As 
a case of GHZ state $\pi_W$ decreases with increasing the temperature, and eventually reduces to $0.18$ at $T_H = \infty$.}
\end{center}
\end{figure}
%%%%%%%%%%%%%%%%%%%%%%%%%%%%%%%%%%%%%%%%%%%%%%%%%%%%%%%%%%%

\section{Conclusion}
In this paper we discussed the quantum discord and tripartite entanglement in the presence of the asymptotically flat static black holes.
Both the quantum discord and the tripartite entanglement exhibit decreasing behavior with increasing Hawking temperature. This implies
that the presence of the black holes reduces the quantum correlation when one party moves from asymptotic to near-horizon regions with
his (or her) own particle detector.

Although we have not commented here, the tripartite entanglement of Alice, Bob, and Charlie's {\it in} state (or AntiCharlie) does not completely
vanish. Probably, this fact implies that some quantum information processes can be performed partially across the black hole horizon. 
To confirm this it seems to be important to compute the teleportation fidelity by making use of the tripartite teleportation scheme\cite{tele-1,tele-2}. 
If the tripartite teleportation is possible, even if incompletely, across the horizon, what this means in the context of causality?
The answer may be important in the context of quantum gravity. We would like to explore this issue in the future.

{\bf Acknowledgement}:
This work was supported by the Kyungnam University Research Fund, 2012.

\end{document}